\documentclass[aps,prd,twocolumn,showpacs,floatfix,
superscriptaddress,amsmath,preprintnumbers]{revtex4}
\usepackage{epsf}              %
\usepackage{latexsym}           %
\usepackage{amscd}              %
\usepackage{amsmath}            
\usepackage{graphicx,subfigure} %
\usepackage{comment}            %
\input xy
\xyoption{all}

\newcommand{\beq}{\begin{equation}}
\newcommand{\eeq}{\end{equation}}
\newcommand{\bea}{\begin{eqnarray}}
\newcommand{\eea}{\end{eqnarray}}

\renewcommand{\o}{\overline}

\newcommand{\benn}{\begin{displaymath}}
\newcommand{\eenn}{\end{displaymath}}

\newcommand{\tw}{\textwidth}
\newcommand{\ig}{\includegraphics}
\def\slashchar#1{\ensuremath{                   %
   \setbox0=\hbox{${}#1{}$}                     
   \dimen0=\wd0                                 
   \setbox1=\hbox{/} \dimen1=\wd1               
   \ifdim\dimen0>\dimen1                        
   \rlap{\hbox to \dimen0{\hfil/\hfil}}         
   {}#1{}                                       
   \else                                        
   \rlap{\hbox to \dimen1{\hfil${}#1{}$\hfil}}  
   /                                            
   \fi}}                                        %
\def\simge{
 \mathrel{\rlap{\raise 0.511ex
  \hbox{$>$}}{\lower 0.511ex \hbox{$\sim$}}}}
\def\simle{
 \mathrel{\rlap{\raise 0.511ex
  \hbox{$<$}}{\lower 0.511ex \hbox{$\sim$}}}}

\begin{document}
 \include{def}
\preprint{LBNL-61613,DOE/ER/40762-375} 

\title{Lattice QCD determination of states with spin $\frac{5}{2}$ or higher in the spectrum of nucleons}

\author{Subhasish Basak}
\affiliation{Department of Physics, N.N.D. College, Calcutta 700092, India}
\author{R.~G.~Edwards}
\affiliation{Thomas Jefferson National Accelerator Facility, Newport News, VA 23606, USA}
\author{G.~T.~Fleming}
\affiliation{Yale University, New Haven, CT 06520, USA}
\author{J.~Juge}
\affiliation{Department of Physics, Carnegie Mellon University, Pittsburgh, PA 15213, USA}
\author{A.~Lichtl}
\affiliation{Department of Physics, Carnegie Mellon University, Pittsburgh, PA 15213, USA}
\author{C.~Morningstar}
\affiliation{Department of Physics, Carnegie Mellon University, Pittsburgh, PA 15213, USA}
\author{D.~G.~Richards}
\affiliation{Thomas Jefferson National Accelerator Facility, Newport News, VA 23606, USA}
\author{I.~Sato}
\affiliation{Lawrence Berkeley Laboratory, Berkeley, CA 94720, USA}
\author{S.~J.~Wallace}
\affiliation{University of Maryland, College Park, MD 20742, USA}

\begin{abstract}
Energies for excited isospin $\frac{1}{2}$
states that include the nucleon are 
computed using quenched, anisotropic lattices. 
Baryon interpolating field operators that are used include
nonlocal operators that provide $G_2$ irreducible representations of the
octahedral group.  The decomposition of spin $\frac{5}{2}$ or higher states is
realized for the first time in a lattice QCD calculation. 
We observe patterns of degenerate energies in the irreducible 
representations of the octahedral group that correspond to the subduction of
the continuum spin $\frac{5}{2}$ or higher. 
\end{abstract}
\maketitle

The theoretical determination of the spectrum of baryon resonances
from the fundamental quark and gluon degrees of freedom is an
important goal for lattice QCD. To date there have been
studies of experimental ground state energies for different
baryons~\cite{Aoki:1999yr,Davies:2003ik,AliKhan:2001tx}
but only a few results for excited state energies have been
reported.~\cite{Zhou:2006xe,Burch:2006cc,Sasaki:2005uq,
Sasaki:2005ug,Basak:2004hr,Zanotti:2003fx} No clear determination 
of states with spin $\frac{5}{2}$ or higher has been reported
because nonlocal operators have not been used. In this work we
find degenerate energies that occur in irreducible representations of
the octahedral group corresponding to the subduction of
the continuum spin $\frac{5}{2}$ or higher. This provides the 
first lattice QCD calculation that realizes the decomposition 
of spins greater than $\frac{3}{2}$.

Lattice correlation functions do not correspond to
definite values of total angular momentum. However, they
do correspond to definite 
irreducible representations (irreps) of the octahedral group when 
the source and sink operators transform accordingly.
There are six double-valued irreps of the octahedral group: three for
even-parity that are labeled with a $g$ subscript ({\it gerade})
and three for odd-parity that are labeled with a $u$ subscript
({\it ungerade}). They are: $G_{1g}, H_g, G_{2g}, G_{1u}, H_u$ and $G_{2u}$.

Continuum values of 
total angular momenta are realized in lattice simulations by
patterns of degenerate energies in the continuum limit that match the patterns 
for the subduction of spin $J$ to the
double-valued irreps of the octahedral group. These 
patterns are shown in Table~\ref{table:subduction}. 
\begin{table}
\caption{The number of occurrences
of double-valued irrep $\Lambda$ of the octahedral group for
different values of continuum spin $J$. 
} \label{table:subduction}
\begin{tabular}{|cllllllll|}
\hline
$\Lambda$   &  &J= & ~${1\over 2}$~&~${3\over 2}$~&~ ${5\over 2}$ ~&~ ${7\over 2}$~&~ ${9\over 2}$~&~ ${11 \over 2}$ \\
$G_{1}$ ~&~  ~&~  ~&~ 1 ~&~ 0 ~&~ 0 ~&~ 1 ~&~ 1 ~&~ 1     \\
$H $    ~&~  ~&~  ~&~ 0 ~&~ 1 ~&~ 1 ~&~ 1 ~&~ 2 ~&~ 2  \\
$G_{2}$ ~&~  ~&~  ~&~ 0 ~&~ 0 ~&~ 1 ~&~ 1 ~&~ 0 ~&~ 1   \\
\hline
\end{tabular}
\end{table}
For example, a state in one of the $G_2$ irreps is a signal for
the subduction of continuum spin $\frac{5}{2}$ or higher.
For spin $\frac{5}{2}$, there must be partner states in the $H$
and $G_2$ irreps that would be degenerate in the continuum limit.  
For spin $\frac{7}{2}$, there must be 
partner states in the $G_1$, $H$ 
and $G_2$ irreps and for spin $\frac{9}{2}$ there must be one 
partner in the $G_1$ irrep and two in the $H$ irrep.  
The partner states should be degenerate in the continuum limit where lattice spacing 
$a \rightarrow 0$.

This paper reports on work to determine
the pattern of low-lying states in the isospin $\frac{1}{2}$
spectrum. We carry out an analysis in quenched lattice QCD using 
a moderately large number of three-quark operators. 
Work is in progress to use a very large number of
operators.\cite{Lichtl-Thesis}
Smeared quark and gluon fields are used.~\cite{Albanese87,Alford:1995dm}
Smearing reduces the  
couplings to short wavelength fluctuations of the theory and
provides cleaner determinations of effective energies. 
This is important when a large array of interpolating 
field operators is used in order to implement the correlation-matrix  method
of Refs.~\cite{michael85,lw90}.

In order to determine suitable operators, including the
nonlocal ones that are required for the $G_2$ irreps, we
have developed sets of
baryon operators that transform according to irreducible representations
using an analytical method based on
appropriate Clebsch-Gordan coefficients for the octahedral group.\cite{Basak:2005ir}
 An alternative, automated procedure that is developed 
in Ref.~\cite{Basak:2005aq}
provides very large sets of operators. Both
 methods provide equivalent results.  
In this work, we use for positive parity the three-quark operators 
defined in Tables VI, VII, and X
of Ref.~\cite{Basak:2005ir}.  These comprise a complete set
of quasilocal operators plus the simplest set of nonlocal
operators that have one quark displaced relative to the other two
quarks.

Negative-parity operators are obtained by applying the 
charge-conjugation transformation to the positive-parity operators.  
A three-quark operator that transforms according to irrep $\Lambda$
and row $\lambda$ of the octahedral group is related by charge conjugation to an
operator that transforms according to irrep $\Lambda_c$ and row $\lambda_c$
and has opposite
parity, i.e.,  
\begin{equation}
 {\cal C} \o B_k^{(\Lambda\lambda)} {\cal C}^\dag
= -B^{(\Lambda_c \lambda_c )}_{k},~~~ 
{\cal C} B_k^{(\Lambda\lambda)} {\cal C}^\dag
 = -\o B^{(\Lambda_c\lambda_c)*}_{k} .
\label{eq:baryon_charge_conj}
\end{equation}
If $\Lambda$ is $G_{1g}, G_{2g}$ or $H_g$, then 
$\Lambda_c$ is $G_{1u}, G_{2u}$ or $H_u$, respectively, and 
vice versa. The row labels are related by $\lambda_c=d_\Lambda +1 - \lambda$, where $d_{\Lambda}$ is the dimension
of the irrep: $d_{G_1} = 2, d_{H}= 4, d_{G_2}=2$.

A Hermitian matrix of correlation
functions for irrep $\Lambda$ and row $\lambda$ is obtained by including
a $\gamma_4$ matrix for each quark field in the 
source operator $\o B^{(\Lambda\lambda)}_{k'}({\bf 0},0)$.  This gives
\begin{equation}
C^{(\Lambda\lambda)}_{kk'}(t)= {\cal P}^{(\Lambda)}_{k'} \sum_{\bf x}
\langle 0 \vert  T \Bigr( B^{(\Lambda\lambda)}_{k}({\bf x},t)
\o B^{(\Lambda\lambda)}_{k'}({\bf 0},0)\Bigr) \vert 0 \rangle,
\label{eq:matrix_correlation_function2}
\end{equation}
where the factor ${\cal P}^{(\Lambda)}_{k'}$ is $\pm1$ and it  
arises from evaluating the action of the $\gamma_4$ matrices
in the source operator. Subscripts $k$ and $k'$ refer to different embeddings
of irrep $\Lambda$.  
The relation between matrices of correlation functions with different parities
is~\cite{Basak:2005aq},
\begin{equation}
C^{(\Lambda\lambda)}_{kk'}(t) = - \eta_t C^{(\Lambda_c \lambda_c)*}_{kk'}(T-t).
\label{eq:PCT}
\end{equation}
where $\eta_t = 1$ for periodic boundary conditions and 
 $\eta_t = -1$ for anti-periodic boundary conditions.
Applying charge conjugation and time-reversal to a correlation
function produces one for the opposite parity because of PCT symmetry.
We use a sufficiently long time extent, $T$, 
so that this symmetry provides two independent correlation functions for 
each parity.

 
Analyses of excited state energies are based upon the matrices of
correlation functions of
Eq.~(\ref{eq:matrix_correlation_function2}).
In order to extract the spectrum of energies from the matrix of
correlation functions, we first average over rows because they
provide equivalent results owing to octahedral symmetry.  We then solve the following
generalized eigenvalue equation,
\begin{equation}
\sum_{k'} \widetilde{C}^{(\Lambda)}_{kk'}(t) v^{(n)}_{k'}
(t,t_0) = \alpha^{(n)}(t,t_0) \sum_{k'}
\widetilde{C}^{(\Lambda)}_{kk'}(t_0) v^{(n)}_{k'}(t,t_0),
\label{eq:new_generalized_eigenvalue_equation}
\end{equation}
where superscript $n$ labels the eigenstates.  The symbol $\widetilde{C}^{(\Lambda)}$
indicates that the appropriate average over rows has been performed. 
The reference time $t_0$ in
Eq.~(\ref{eq:new_generalized_eigenvalue_equation}) is taken near the
source time $t=0$ in order to have significant contributions from
excited states. The principal eigenvalues $\alpha^{(n)}(t,t_0)$
are related to the energy $E_n$ by~\cite{Luscher:1990ck}
\begin{equation}
\alpha^{(n)}(t,t_0) \simeq e^{-E_n(t-t_0)}\Big( 1 + {\cal O}
( e^{-|\delta
E|t}) \Big), \label{eq:generalized_eigenvalue}
\end{equation}
where $\delta E$ is the difference between $E_n$ and the next
closest energy.
 We have determined effective energies $E_n$ by fitting the
generalized eigenvalues to the leading term of Eq.~(\ref{eq:generalized_eigenvalue}).  

Methods that improve the signal-to-noise ratio
are very important for extracting excited state energies from
lattice QCD simulations.
The correlation-matrix method optimizes the operators 
so as to achieve early plateaus of
the effective energies and the use of PCT 
symmetry increases the statistics. In addition, we use 
anisotropic lattices with temporal lattice
spacing $a_t$ one-third of the spatial lattice spacing $a_s$. The
finer spacing increases the number of
time-slices for extraction of energies before the signal for 
a high-energy state decays to the
level of the noise.   

In this work anisotropic lattices with two
different volumes are used: 239
gauge field configurations are used for a $16^3\times 64$ lattice
and 167 configurations are used for a $24^3\times 64$ lattice.
Because of the use of PCT symmetry, the size of the statistical ensemble
is effectively doubled.  
Gauge-field configurations are generated using the anisotropic,
unimproved Wilson gauge action
\cite{Chen:2000ej,Klassen:1998ua,Karsch:1997dc,Fujisaki:1996vv}
 in the quenched approximation
with $\beta = 6.1$ and the pion mass is 490 MeV.  
For both lattices, the temporal lattice spacing
corresponds to $a^{-1}_t = 6.05(1)$ Gev~\cite{Basak:2004hr} as determined
from the string tension.  Calculations are performed using the
Chroma software.~\cite{Chroma}
Further details of the action and lattice parameters will be presented elsewhere.

We have extracted energies for isospin $\frac{1}{2}$ and $\frac{3}{2}$
channels by
diagonalizing matrices of correlation functions formed from
three-quark operators that share the same octahedral 
symmetry. Starting with a large number
of operators, we first eliminate operators that 
have less influence on the determination of the effective energies and 
the eigenvectors ${\bf v}^{(n)}$ of the low-lying states.  
Once a set of good operators is obtained, we form final
matrices of correlation functions, diagonalize them and extract
effective energies. 
This procedure yields solid results for energies of low-lying states 
and is more efficient
than diagonalizing matrices of the
largest dimension. 
We have obtained 17 energies for nucleonic states and 11
energies for delta baryon states.  In this paper, we focus 
on the nucleon channel and the determination of spin. 



 
\begin{figure}[h]
\hspace{0cm}
\ig[width=0.48\tw]{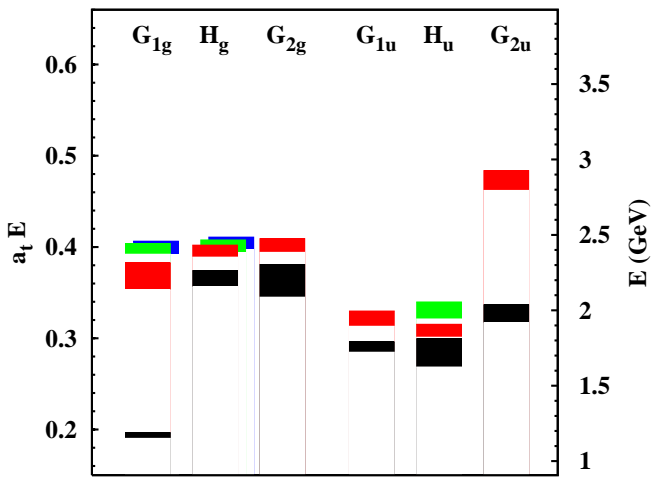} 
\caption{The energies obtained for each symmetry channel 
of isospin ${1\over 2}$ baryons are shown 
based on the $24^3\times64$ lattice data.
The scale on the left side shows energies in lattice
units and the scale on the right side shows energies in GeV. Errors are indicated 
by the vertical size of the box.} 
\label{fig:N24energies}
\end{figure}

Our lattice results for the spectrum are shown in Fig.~\ref{fig:N24energies}.
For positive parity, the nucleon state is very well determined
at lattice energy $a_tE = 0.193(3)$.  It is an isolated 
$G_{1g}$ state corresponding to spin $\frac{1}{2}$. 
Well above the nucleon is a group of three degenerate 
states (within errors) in the 
$G_{1g}$, $H_g$ and $G_{2g}$ irreps. 
The results on both lattice volumes are essentially the same, 
although the lowest $G_{2g}$ state in the smaller volume has a larger error.

Energies of $G_2$ states are the
most difficult to extract because the
plateaus are relatively short-lived, signal-to-noise ratios are
large, and the number of operators that we have used for the
matrices of correlation functions is limited. 
 The limitation on the number of
operators is because we restrict the operators used in this work to ones that
can be constructed from one-link displacements. In work that is in progress,
more varied types of operators are used to obtain very larger numbers of 
$G_2$ operators, which is essential for determining states with higher
energies.

\begin{figure}[h]
\hspace{0cm}
\ig[width=0.48\tw]{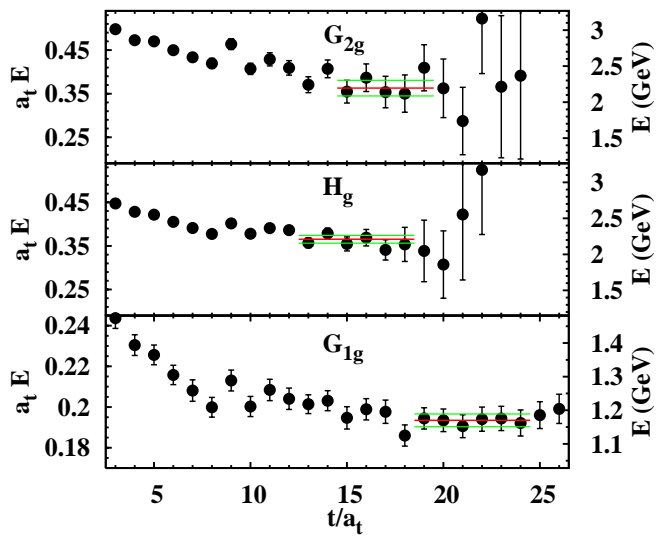} 
\caption{Effective energies for the lowest states in $G_{1g}, H_{g}$ and 
$G_{2g}$
irreps.  Time ranges used for and energies obtained from fits to the principal
eigenvalues are shown by horizontal lines. } 
\label{fig:EffMass_N24}
\end{figure}
 Figure~\ref{fig:EffMass_N24}
shows effective masses obtained for the lowest energy $G_{1g}$, $H_g$ and
$G_{2g}$ states. The horizontal lines indicate the time ranges used to
fit the principal eigenvalues $\alpha^{(n)}(t,t_0)$ by a single exponential form.
The resulting mean energy value and the error range are indicated by the lines.

%

Although there are substantial discretization errors with the quark 
action that is used, and they could contribute differently in 
the different irreps, clear patterns in the degeneracies emerge.
Focusing on the group of three positive-parity states 
near lattice energy $a_t E = 0.36$ in Fig.~\ref{fig:N24energies},
two interpretations are possible. 
A.) The group consists of a spin $\frac{1}{2}$ state and a 
spin $\frac{5}{2}$ state that are accidentally
degenerate.  In this case the $G_{1g}$ state corresponds to spin 
$\frac{1}{2}$ and the $H_g$ and $G_{2g}$ partner states
correspond to the subduction of spin $\frac{5}{2}$. 
B.) The group consists of a single state
with the degenerate $G_{1g},~ H_g$ and $G_{2g}$ 
partner states corresponding to the subduction of spin $\frac{7}{2}$.
 
In the physical spectrum of positive-parity nucleon resonances, 
the lowest excited state, $N(1440,\frac{1}{2}^+)$, lies below all
negative parity states. We do not find a signal for a positive-parity 
excitation that has lower energy than the negative-parity excitations
at this quark mass.
The next two 
excited nucleon states are essentially degenerate,
namely, $N(1680,\frac{5}{2}^+)$ and 
$N(1710,\frac{1}{2}^+)$, each with a width of about 100 MeV.  
Spin $\frac{7}{2}$ states occur only at significantly higher energy
(1990 MeV).  
Primarily because of the absence of spin $\frac{7}{2}$ in the low-lying
spectrum,  interpretation A.) of our lattice results is more
consistent with the physical pattern of energies and spins.
In the absence of $G_2$ operators, previous lattice studies have
assumed that states obtained with 
$H$ irrep operators correspond to continuum spin 
$\frac{3}{2}$.~\cite{Zhou:2006xe,Sasaki:2005uq,Sasaki:2005ug,Zanotti:2003fx}.
With the $G_2$ operators, we find low-lying states 
in irrep $H$ that are consistent with the subduction 
of spin $\frac{5}{2}$ or higher.  

In the negative-parity spectrum, we also obtain essentially the
same results for both lattice volumes. The three lowest states shown 
on the right half of Fig.~\ref{fig:N24energies} are unambiguously identified
as follows: the lowest
$G_{1u}$ state corresponds to spin $\frac{1}{2}$ and the lowest two
$H_u$ states correspond to distinct spin $\frac{3}{2}$ states.
Above these is a group of three states with roughly the same 
lattice energy: $a_t E \approx 0.33$ (within errors).
Again there are two possible interpretations.  C.) The group consists of 
a spin $\frac{1}{2}$ state in $G_{1u}$ that is accidentally degenerate 
with a spin $\frac{5}{2}$ state, the latter having degenerate partner states 
in $H_u$ and $G_{2u}$.  D.) The group consists a spin $\frac{7}{2}$ 
state having degenerate partner states in $G_{1u}$, $H_u$ and $G_{2u}$.  

The pattern of low-lying physical states starts with
$N(\frac{3}{2}^-,1520)$, $N(\frac{1}{2}^-,1535)$
and  $N(\frac{3}{2}^-,1700)$. These should show up
as distinct $H_u$, $G_{1u}$ and $H_u$ states on the lattice, in agreement with
the three lowest negative-parity states in Fig.~\ref{fig:N24energies}.
The next physical states include $N(\frac{1}{2}^-,1680)$ and $N(\frac{5}{2}^-,1675)$,
which essentially are degenerate.  They should show up as
degenerate $G_{1u}$, $H_u$ and $G_{2u}$ states on the lattice.  
This pattern of spins is consistent with interpretation C.) of the lattice 
states at lattice energy $a_t E \approx 0.33$.
The pattern of energies of the physical states has $N(\frac{5}{2}^-,1675)$
a little lower in energy than $N(\frac{3}{2}^-,1700)$, but the
lattice results at lattice spacing $0.1 F$ place the 
spin $\frac{5}{2}$ state above the spin $\frac{3}{2}$ state.  
Study of the continuum limit of the lattice spectrum is required
in order to resolve these issues.  

Because the minimum spin that is contained in the $G_2$ irrep 
is $\frac{5}{2}$, we have found strong evidence for spin
$\frac{5}{2}$ or higher for both parities in our lattice spectra for
isospin $\frac{1}{2}$.    
We also have found evidence for degenerate
partner states corresponding to the subduction 
of spin $\frac{5}{2}$ or higher to the octahedral irreps. 
The lattice results for the 
low-lying excited states of isospin $\frac{1}{2}$ 
provide the correct number of octahedral 
states for the subduction of the spins of the low-lying physical states. 
These results are significant but they are based on the quenched 
approximation and a 490 MeV pion mass.  Similar calculations in full QCD
with several pion masses and several lattice spacings are planned
and work is under way to calculate the anisotropic gauge configurations.
\acknowledgements
This work was supported by the U.S. National Science Foundation under
Award PHY-0354982 and by the U.S. Department of Energy
under contracts DE-AC05-06OR23177 and DE-FG02-93ER-40762.


\begin{thebibliography}{10}

\bibitem{Aoki:1999yr}
  S.~Aoki {\it et al.}  [CP-PACS Collaboration],
  Phys.\ Rev.\ Lett.\  {\bf 84}, 238 (2000)
  [arXiv:hep-lat/9904012].

\bibitem{Davies:2003ik}
  C.~T.~H.~Davies {\it et al.}  [HPQCD Collaboration],
  Phys.\ Rev.\ Lett.\  {\bf 92}, 022001 (2004)
  [arXiv:hep-lat/0304004].

\bibitem{AliKhan:2001tx}
  A.~Ali Khan {\it et al.}  [CP-PACS Collaboration],
  Phys.\ Rev.\ D {\bf 65}, 054505 (2002)
  [Erratum-ibid.\ D {\bf 67}, 059901 (2003)]
  [arXiv:hep-lat/0105015].

\bibitem{Zhou:2006xe}
  L.~Zhou and F.~X.~Lee,
  arXiv:hep-lat/0604023.

\bibitem{Burch:2006cc}
  T.~Burch, C.~Gattringer, L.~Y.~Glozman, C.~Hagen, D.~Hierl, 
  C.~B.~Lang and A.~Schafer,
  arXiv:hep-lat/0604019.

\bibitem{Sasaki:2005uq}
  K.~Sasaki and S.~Sasaki,
  PoS {\bf LAT2005}, 060 (2005)
  [arXiv:hep-lat/0508026].

\bibitem{Sasaki:2005ug}
  K.~Sasaki and S.~Sasaki,
  Phys.\ Rev.\ D {\bf 72}, 034502 (2005)
  [arXiv:hep-lat/0503026].


\bibitem{Basak:2004hr}
  S.~Basak {\it et al.}  [LHP Collaboration],
  Nucl.\ Phys.\ Proc.\ Suppl.\  {\bf 140}, 278 (2005)
  [arXiv:hep-lat/0409082].

\bibitem{Zanotti:2003fx}
  J.~M.~Zanotti, D.~B.~Leinweber, A.~G.~Williams, J.~B.~Zhang, 
  W.~Melnitchouk and S.~Choe
                  [CSSM Lattice collaboration],
  Phys.\ Rev.\ D {\bf 68}, 054506 (2003)
  [arXiv:hep-lat/0304001].

\bibitem{Albanese87}
M.~Albanese \textit{et al.} (APE Collaboration),
Phys.\ Lett.\ B{\bf 192}, 163 (1987).

\bibitem{Alford:1995dm}
  M.~G.~Alford, T.~Klassen and P.~Lepage,
  Nucl.\ Phys.\ Proc.\ Suppl.\  {\bf 47}, 370 (1996)
  [arXiv:hep-lat/9509087].

\bibitem{Lichtl-Thesis}
  A.~Lichtl, [arXiv:hep-lat/0609019]. 
  
\bibitem{michael85}
  C.~Michael, Nucl.\ Phys.\ B{\bf 259}, 58 (1985).

\bibitem{lw90}
  M.~L\"{u}scher and U.~Wolff, Nucl.\ Phys.\ B{\bf 339}, 222 (1990).

\bibitem{Basak:2005ir}
  S.~Basak {\it et al.}  [Lattice Hadron Physics Collaboration (LHPC)],
  Phys.\ Rev.\ D {\bf 72}, 074501 (2005)
  [arXiv:hep-lat/0508018].

\bibitem{Basak:2005aq}
  S.~Basak {\it et al.},
  Phys.\ Rev.\ D {\bf 72}, 094506 (2005)
  [arXiv:hep-lat/0506029].

\bibitem{Luscher:1990ck}
  M.~Luscher and U.~Wolff,
  Nucl.\ Phys.\ B {\bf 339}, 222 (1990).

\bibitem{Chen:2000ej}
  P.~Chen,
  Phys.\ Rev.\ D {\bf 64}, 034509 (2001)
  [arXiv:hep-lat/0006019].

\bibitem{Klassen:1998ua}
  T.~R.~Klassen,
  Nucl.\ Phys.\ B {\bf 533}, 557 (1998)
  [arXiv:hep-lat/9803010].

\bibitem{Karsch:1997dc}
  F.~Karsch, J.~Engels and T.~Scheideler,
  Nucl.\ Phys.\ Proc.\ Suppl.\  {\bf 63}, 427 (1998)
  [arXiv:hep-lat/9709011].

\bibitem{Fujisaki:1996vv}
  M.~Fujisaki {\it et al.}  [QCD-TARO Collaboration],
  Nucl.\ Phys.\ Proc.\ Suppl.\  {\bf 53}, 426 (1997)
  [arXiv:hep-lat/9609021].

\bibitem{Chroma}
R.~G.~Edwards (LHPC Collaboration)and B.~Joo (UKQCD Collaboration), "The Chroma
Software System for Lattice QCD", arXiv:hep-lat/0409003, Proceedings of the 22nd
International Symposium for Lattice Field Theory (Lattice2004), Nucl. Phys B140
(Proc. Suppl) p832, 2005.

\end{thebibliography}
\end{document}